\documentstyle[12pt]{article}
\newcommand{\eq}{\begin{equation}}
\newcommand{\en}{\end{equation}}
\newcommand{\eqn}{\begin{eqnarray}}
\newcommand{\enn}{\end{eqnarray}}
\begin{document}
\begin{titlepage}
\begin{flushright}
hep-th/9403007 \\
THEP-94-2 \\
2 March 1994
\end{flushright}
\begin{center}
\LARGE
Higgs and Fermions in $D_{4}-D_{5}-E_{6}$ Model \\
based on $Cl(0,8)$ Clifford Algebra \\
\vspace{0.5cm}
\large
 Frank D. (Tony) Smith, Jr. \\
\vspace{6pt}

\small
Department of Physics \\
Georgia Institute of Technology \\
Atlanta, Georgia 30332 \\
\vspace{0.5cm}
{\bf Abstract}
\end{center}

In the $D_{4}-D_{5}-E_{6}$ model of a series of papers
(hep-ph/9301210, \\
hep-th/9302030, hep-th/9306011,
and hep-th/9402003) \cite{SM1, SM2, SM3, SM4} \\
an 8-dimensional spacetime with Lagrangian action  \\
\[
\int_{V_{8}} F_{8} \wedge \star F_{8} + \partial_{8}^{2}
\overline{\Phi_{8}} \wedge \star \partial_{8}^{2} \Phi_{8} +
\overline{S_{8\pm}} \not \! \partial_{8} S_{8\pm} +
GF + GH
\]
is reduced to a 4-dimensional Lagrangian. \\
In \cite{SM4}, the gauge boson terms were seen to give \\
$SU(3) \times SU(2) \times U(1)$ for the color, weak,
and electromagnetic forces and \\
gravity of the MacDowell-Mansouri type \cite{McM}, \\
which has recently been shown by Nieto, Obreg\'{o}n,
and Socorro \cite{NOS} \\
in gr-qc/9402029 to be equivalent, up to a
Pontrjagin topological term, \\
to the Ashtekar formulation. \\

This paper discusses the Higgs and spinor fermion terms.

\vspace{12pt}
\normalsize
\footnoterule
\noindent
{\footnotesize \copyright 1994 Frank D. (Tony) Smith, Jr.,
341 Blanton Road, Atlanta, Georgia 30342 USA \\
P. O. Box for snail-mail:    P. O. Box 430, Cartersville,
Georgia 30120 USA \\
e-mail:    gt0109e@prism.gatech.edu
and fsmith@pinet.aip.org} \\
WWW URL:    http://www.gatech.edu/tsmith/home.html
\end{titlepage}

\newpage

\setcounter{footnote}{0}
\section{Introduction}
\setcounter{equation}{0}

	The $D_{4}-D_{5}-E_{6}$ model of physics starts out with
an 8-dimensional spacetime that is reduced to a 4-dimensional
spacetime.

The 8-dimensional Lagrangian (up to gauge-fixing and ghost
terms) is:

$$
\int_{V_{8}} F_{8} \wedge \star F_{8} + \partial_{8}^{2}
\overline{\Phi_{8}} \wedge \star \partial_{8}^{2} \Phi_{8} +
\overline{S_{8\pm}} \not \! \partial_{8} S_{8\pm} +
GF + GH
$$

where $F_{8}$ is the 28-dimensional $Spin(8)$ curvature,
$\star$ is the Hodge dual, \\
$\partial_{8}$ is the 8-dimensional covariant derivative, \\
$\Phi_{8}$ is the 8-dimensional scalar field, \\
$\not \!  \partial_{8}$ is the 8-dimensional Dirac operator, \\
$V_{8}$ is 8-dimensional spacetime, \\
$S_{8\pm}$ are the $+$ and $-$ 8-dimensional
half-spinor fermion spaces, and \\
GF and GH are gauge-fixing and ghost terms.

\vspace{12pt}

(hep-th/9402003 \cite{SM4} had a typo error of
$S_{8+}$ or $S_{8-}$ instead of $S_{8\pm}$.)

\vspace{12pt}

This paper describes the Higgs mechanism and the spinor
fermions of the 4-dimensional Lagrangian.
Results of the preceding papers in this series
\cite{SM1, SM2, SM3, SM4} are assumed.  They are
hep-ph/9301210, hep-th/9302030, hep-th/9306011, and
hep-th/9402003).

\vspace{12pt}

{\bf Table of Contents:}

2.  Scalar part of the Lagrangian

2.1  First term $F_{H44} \wedge \star F_{H44}$

2.2  Third term $\int_{\perp 4} F_{H\perp 4 \perp 4}
\wedge \star F_{H\perp 4 \perp 4}$

2.3  Second term $F_{H4 \perp 4} \wedge \star F_{H4 \perp 4}$

2.4  Higgs Mass

3.  Spinor Fermion part of the Lagrangian

3.1  Yukawa Coupling and Fermion Masses

4.  Parity Violation, W-Boson Masses, and $\theta_{w}$

4.1  Massless Neutrinos and Parity Violation

4.2  $W_{0}$, $Z$, and $\theta_{w}$

4.3  Corrections for $m_{Z}$ and $\theta_{w}$

5.  Some Errata for Previous Papers

\vspace{12pt}

\newpage

{\bf Summary of Some Material from Earlier Papers:}

\vspace{12pt}

$S_{8\pm} = {\bf{R}}P^{1} \times S^{7}$ \cite{SM2});

\vspace{12pt}

$S_{8\pm} \oplus S_{8\pm} = ({\bf{R}}P^{1} \times S^{7})
\oplus ({\bf{R}}P^{1} \times S^{7})$ is the full fermion
space of first generation particles and antiparticles,
and is the Silov boundary of the 32(real)-dimensional
bounded complex domain corresponding to the $Type V$ HJTS
$E_{6}/(Spin(10) \times U(1)$ \cite{SM2}

(hep-th/9302030 \cite{SM2} had erroneously
used $\times$ instead of $\oplus$.);

\vspace{12pt}

after dimensional reduction, the weak force gauge
group is $SU(2)$ \cite{SM4};

\vspace{12pt}

with respect to $SU(2)$ of the Higgs and weak force,
the 4-dimensional spacetime manifold has global
type $M = S^{2} \times S^{2}$ \cite{SM2};

\vspace{12pt}

the Higgs and weak force $SU(2)$ acts effectively on a
submanifold of the half-spinor fermion space
$S_{8\pm} = {\bf{R}}P^{1} \times S^{7}$, that is,
$Q_{3} = {\bf{R}}P^{1} \times S^{2}$, which is Silov boundary
of the 6(real)-dimensional bounded complex domain
corresponding to the $Type IV_{3}$ HJTS
$\overline{D_{3}} = Spin(5)/(SU(2) \times U(1)$ \cite{SM2}

(hep-th/9302030 \cite{SM2} had erroneously listed
$SU(3)/SU(2) \times U(1)$ instead of
$Spin(5)/SU(2) \times U(1) = Spin(5)/Spin(3) \times U(1)$.);

\vspace{12pt}

after dimensional reduction, the Higgs scalar is the
4th component of the column minimal ideal of $Cle(0,6)$
that contains $W_{+}$, $W_{-}$, and $W_{0}$ of the
$SU(2)$ weak force, and so is an $SU(2)$ scalar \cite{SM4};

the $SU(2)$ gauge group of the vector bosons $W_{+}$, $W_{-}$,
and $W_{0}$ and the $SU(2)$ gauge group of the scalar Higgs,
insofar as they are independent,
can be considered as one $Spin(4) = SU(2) \times
SU(2)$ gauge group \cite{SM2, SM4};
and

\vspace{12pt}

the electromagnetic $U(1)$ of the $D_{4}-D_{5}-E_{6}$ model
comes from the $U(3)$ containing the color force $SU(3)$,
and so is "unified" with the $SU(3)$ color force rather
than with the $SU(2)$ weak force \cite{SM4}.

\vspace{12pt}

The last statement is different from most formulations of
the standard model, but is similar to the formulation of
O'Raifeartaigh (section 9.4 of \cite{LO})
of the standard model as $S(U(3) \times U(2))$ rather
than $SU(3) \times U(2)$ or $SU(3) \times SU(2) \times U(1)$.
O'Raifeartaigh states that the unbroken gauge
symmetry is actually $U(3)$ rather than $SU(3) \times U(1)$.

\pagebreak

\section{Scalar part of the Lagrangian}

The scalar part of the 8-dimensional Lagrangian is

$$
\int_{V_{8}}
\partial_{8}^{2} \overline{\Phi_{8}} \wedge \star
\partial_{8}^{2} \Phi_{8}
$$

As shown in chapter 4 of G\"{o}ckeler and Sch\"{u}cker
\cite{GS}, $\partial_{8}^{2} \Phi_{8}$ can be represented
as an 8-dimensional curvature $F_{H8}$, giving

$$
\int_{V_{8}} F_{H8} \wedge \star F_{H8}
$$

When spacetime is reduced to 4 dimensions, denote the
surviving 4 dimensions by $4$ and the reduced
4 dimensions by $\perp 4$.

\vspace{12pt}

Then, $F_{H8} = F_{H44} + F_{H4\perp 4} + F_{H\perp 4 \perp 4}$,
where

$F_{H44}$ is the part of $F_{H8}$ entirely in the
surviving spacetime;

$F_{H4 \perp 4}$ is the part of $F_{H8}$ partly in the
surviving spacetime and partly in the reduced spacetime; and

$F_{H\perp 4 \perp 4}$ is the part of $F_{H8}$ entirely in the
reduced spacetime;

\vspace{12pt}

The 4-dimensional Higgs Lagrangian is then:

$\int (F_{H44} + F_{H4 \perp 4} + F_{H\perp 4 \perp 4})
\wedge
\star (F_{H44} + F_{H4 \perp 4} + F_{H\perp 4 \perp 4}) =$

$=\int (F_{H44} \wedge \star F_{H44} +
F_{H4 \perp 4} \wedge \star F_{H4 \perp 4} +
F_{H\perp 4 \perp 4} \wedge \star F_{H\perp 4 \perp 4})$.

\vspace{12pt}

As all possible paths should be taken into account in
the sum over histories path integral picture of quantum
field theory, the terms involving the reduced 4 dimensions,
$\perp 4$, should be integrated over the reduced 4 dimensions.

\vspace{12pt}

Integrating over the reduced 4 dimensions, $\perp 4$, gives

$\int \left(  F_{H44} \wedge \star F_{H44}  +
 \int_{\perp 4} F_{H4 \perp 4} \wedge \star F_{H4 \perp 4} +
\int_{\perp 4} F_{H\perp 4 \perp 4} \wedge
\star F_{H\perp 4 \perp 4} \right)$.

\vspace{12pt}

\subsection{First term $ F_{H44} \wedge \star F_{H44}$}

\vspace{12pt}

The first term is just $\int F_{H44} \wedge \star F_{H44}$.

Since they are both $SU(2)$ gauge boson terms, this term,
in 4-dimensional spacetime, just merges into the $SU(2)$ weak
force term $\int F_{w} \wedge \star F_{w}$.

\vspace{12pt}

\subsection{Third term $\int_{\perp 4} F_{H\perp 4 \perp 4}
\wedge \star F_{H\perp 4 \perp 4}$}

\vspace{12pt}

The third term,
$ \int \int_{\perp 4} F_{H\perp 4 \perp 4} \wedge
\star F_{H\perp 4 \perp 4}$,
after integration over $\perp 4$,
produces terms of the form

$ \lambda (\overline{\Phi} \Phi)^{2} - \mu^{2}
\overline{\Phi} \Phi$
by a process similar to the Mayer mechanism.

\vspace{12pt}

The Mayer mechanism is based on Proposition 11.4 of

chapter 11 of volume I of Kobayashi and Nomizu \cite{KN},
stating that:

$2 F_{H\perp 4 \perp 4}(X,Y) = [\Lambda(X), \Lambda(Y)] -
\Lambda([X,Y])$,

where $\Lambda$ takes values in the $SU(2)$ Lie algebra.

\vspace{12pt}

If the action of the Hodge dual $\star$ on $\Lambda$ is
such that

$\star \Lambda = - \Lambda$ and $\star [\Lambda, \Lambda] =
[\Lambda, \Lambda]$,

then

$F_{H\perp 4 \perp 4}(X,Y) \wedge
\star F_{H\perp 4 \perp 4}(X,Y) =
(1/4)([\Lambda(X), \Lambda(Y)]^{2} - \Lambda([X,Y])^{2} )$.

\vspace{12pt}

If integration of $\Lambda$ over $\perp 4$ is
$\int_{\perp 4} \Lambda \propto \Phi = (\Phi^{+}, \Phi^{0})$,
then

\vspace{12pt}

$\int_{\perp 4} F_{H\perp 4 \perp 4} \wedge
\star F_{H\perp 4 \perp 4} = $
$ (1/4) \int_{\perp 4} [\Lambda(X),\Lambda(Y)]^{2} -
\Lambda([X,Y])^{2} = $

\vspace{12pt}

$= (1/4) [ \lambda ( \overline{\Phi} \Phi)^{2} - \mu^{2}
\overline{\Phi} \Phi ]$,

\vspace{12pt}

where $\lambda$ is the strength of the scalar field
self-interaction,
$\mu^{2}$ is the other constant in the Higgs potential, and
where $\Phi$ is a 0-form taking values in
the $SU(2)$ Lie algebra.

\vspace{12pt}

The $SU(2)$ values of $\Phi$ are represented by complex

$SU(2) = Spin(3)$ doublets $\Phi = (\Phi^{+}, \Phi^{0})$.

\vspace{12pt}

In real terms,
 $\Phi^{+} = (\Phi_{1} + i \Phi_{2})/ \sqrt{2}$
and
$\Phi^{0} = (\Phi_{3} + i \Phi_{4})/ \sqrt{2}$,

so $\Phi$ has 4 real degrees of freedom.

\vspace{12pt}

In terms of real components,
$\overline{\Phi} \Phi = (\Phi_{1}^{2} + \Phi_{2}^{2} +
\Phi_{3}^{2} + \Phi_{4}^{2})/2 $.

\vspace{12pt}

The nonzero vacuum expectation value of the

$ \lambda (\overline{\Phi} \Phi)^{2} - \mu^{2}
\overline{\Phi} \Phi$ term
is $v = \mu / \sqrt{\lambda}$, and

$<\Phi^{0}> = <\Phi_{3}> = v / \sqrt{2}$.

\vspace{12pt}

In the unitary gauge, $\Phi_{1} = \Phi_{2} = \Phi_{4} = 0$,

and

$\Phi = (\Phi^{+}, \Phi^{0}) = (1/ \sqrt{2})(\Phi_{1} + i
\Phi_{2}, \Phi_{3} + i \Phi_{4}) = (1/ \sqrt{2})
(0, v + H)$,

\vspace{12pt}

where $\Phi_{3} = (v + H) / \sqrt{2}$,

$v$ is the Higgs potential vacuum expectation value, and

$H$ is the real surviving Higgs scalar field.

\vspace{12pt}

Since $\lambda = \mu^{2} / v^{2}$ and $\Phi = (v + H)
/ \sqrt{2}$,

\vspace{12pt}

$(1/4)[ \lambda (\overline{\Phi} \Phi)^{2} - \mu^{2}
\overline{\Phi} \Phi ] = $

\vspace{12pt}

$= (1/16) (\mu^{2} / v^{2})(v + H)^{4} -
(1/8) \mu^{2} (v + H)^{2} = $

\vspace{12pt}

$= (1/16) [ \mu^{2} v^{2} + 4 \mu^{2} vH +
6 \mu^{2} H^{2} + 4 \mu^{2} H^{3} / v + \mu^{2} H^{4} /
v^{2} - 2 \mu^{2} v^{2} - $

$- 4 \mu^{2} v H - 2 \mu^{2} H^{2} ] = $

\vspace{12pt}

$= (1/4) \mu^{2} H^{2} - (1/16) \mu^{2} v^{2}
[ 1 - 4 H^{3} / v^{3} - H^{4} / v^{4} ] $.

\vspace{12pt}

\subsection{Second term $F_{H4 \perp 4} \wedge
\star F_{H4 \perp 4}$}

\vspace{12pt}

The second term,

$\int_{\perp 4} F_{H4 \perp 4} \wedge \star F_{H4 \perp 4}$,

gives $\int \partial \overline{\Phi} \partial \Phi$,
by a process similar to the Mayer mechanism.

\vspace{12pt}

{}From Proposition 11.4 of chapter 11 of volume I of Kobayashi
and Nomizu \cite{KN}:

$2 F_{H4 \perp 4}(X,Y) = [\Lambda(X), \Lambda(Y)] -
\Lambda([X,Y])$,

where $\Lambda$ takes values in the $SU(2)$ Lie algebra.

\vspace{12pt}

For example, if the $X$ component of $F_{H4 \perp 4}(X,Y)$
is in the surviving $4$ spacetime and the $Y$ component of
$F_{H4 \perp 4}(X,Y)$ is in $\perp 4$, then

\vspace{12pt}

the Lie bracket product $ [X,Y] = 0$
so that $\Lambda([X,Y]) = 0$
and therefore

$F_{H4 \perp 4}(X,Y) = (1/2) [\Lambda(X),\Lambda(Y)] =
(1/2) \partial_{X} \Lambda(Y) $.

\vspace{12pt}

The total value of $F_{H4 \perp 4}(X,Y)$ is then
$F_{H4 \perp 4}(X,Y) = \partial_{X}\Lambda(Y) $.

\vspace{12pt}

Integration of $\Lambda$ over $\perp 4$ gives

$\int _{Y \epsilon \perp 4}  \partial_{X}\Lambda(Y) =
\partial_{X}\Phi$,

where, as above, $\Phi$ is a 0-form taking values in the
$SU(2)$ Lie algebra.

\vspace{12pt}

As above, the $SU(2)$ values of $\Phi$ are represented by
complex
$SU(2)=Spin(3)$ doublets $\Phi = (\Phi^{+}, \Phi^{0})$.

\vspace{12pt}

In real terms, $\Phi^{+} = (\Phi_{1} + i \Phi_{2}) /
\sqrt{2}$ and
$\Phi^{0} = (\Phi_{3} + i \Phi_{4}) / \sqrt{2}$,

so $\Phi$ has 4 real degrees of freedom.

\vspace{12pt}

As discussed above, in the unitary gauge,
$ \Phi_{1} = \Phi_{2} = \Phi_{4} = 0$, and

$\Phi = (\Phi^{+}, \Phi^{0}) =
(1/ \sqrt{2})(\Phi_{1} + i \Phi_{2},
\Phi_{3} + i \Phi_{4}) = (1 / \sqrt{2})(0, v + H)$,

where $\Phi_{3} = (v + H) / \sqrt{2}$ ,

$v$ is the Higgs potential vacuum expectation value, and

$H$ is the real surviving Higgs scalar field.

\vspace{12pt}

The second term is then:

$\int  (\int_{\perp 4} - F_{H4 \perp 4} \wedge
\star F_{H4 \perp 4}) =$

$= \int (\int_{\perp 4} (-1/2) [\Lambda(X),\Lambda(Y)] \wedge
\star [\Lambda(X),\Lambda(Y)] ) = \int \partial
\overline{\Phi} \wedge \star \partial \Phi$

\vspace{12pt}

where the $SU(2)$ covariant derivative $\partial$ is

$\partial = \partial + \sqrt{\alpha_{w}} (W_{+} +
W_{-}) + \sqrt{\alpha_{w}} \cos{\theta_{w}}^{2} W_{0}$,
and $\theta_{w}$ is the Weinberg angle.

\vspace{12pt}

Then $\partial \Phi = \partial (v + H) /
 \sqrt{2} =$

$= [\partial H + \sqrt{\alpha_{w}} W_{+} (v + H) +
 \sqrt{\alpha_{w}} W_{-} (v + H) + \sqrt{\alpha_{w}} W_{0}
 (v + H) ] / \sqrt{2}$.

\vspace{12pt}

In the $D_{4}-D_{5}-E_{6}$ model the $W_{+}$, $W_{-}$,
 $W_{0}$, and $H$ terms are considered to be linearly
 independent.

\vspace{12pt}

$v = v_{+} + v_{-} + v_{0}$ has linearly
 independent components  $v_{+}$, $v_{-}$, and $v_{0}$ for
 $W_{+}$, $W_{-}$, and $W_{0}$.

\vspace{12pt}

$H$ is the Higgs component.

\vspace{12pt}

$\partial \overline{\Phi} \wedge \star \partial \Phi$ is
the sum of the squares of the individual terms.

\vspace{12pt}

Integration over $\perp 4$ involving two derivatives
$\partial_{X} \partial_{X}$ is taken to
change the sign by $i^{2} = -1$.

\vspace{12pt}

Then:

$\partial \overline{\Phi} \wedge \star \partial \Phi =
(1/2) (\partial H)^{2} +$

$+ (1/2) [ \alpha_{w} v_{+}^{2} \overline{W_{+}} W_{+} +
  \alpha_{w} v_{-}^{2} \overline{W_{-}} W_{-} +
  \alpha_{w} v_{0}^{2} \overline{W_{0}} W_{0} ] +$

$+ (1/2) [ \alpha_{w} \overline{W_{+}} W_{+} +
  \alpha_{w} \overline{W_{-}} W_{-} +
  \alpha_{w} \overline{W_{0}} W_{0} ] [ H^{2} + 2 v H ]$.

\vspace{12pt}

Then the full curvature term of the weak-Higgs Lagrangian,

$\int F_{w} \wedge \star F_{w} +  \partial
\overline{\Phi} \wedge \star \partial \Phi +
\lambda (\overline{\Phi} \Phi)^{2}  -
\mu^{2} \overline{\Phi} \Phi$,

\vspace{12pt}

is, by the Higgs mechanism:

\vspace{12pt}

$\int [ F_{w} \wedge \star F_{w}  +$

$+ (1/2) [ \alpha_{w} v_{+}^{2} \overline{W_{+}} W_{+} +
  \alpha_{w} v_{-}^{2} \overline{W_{-}} W_{-} +
  \alpha_{w} v_{0}^{2} \overline{W_{0}} W_{0} ] +$

$+ (1/2) [  \alpha_{w} \overline{W_{+}} W_{+} +
  \alpha_{w} \overline{W_{-}} W_{-} +
  \alpha_{w} \overline{W_{0}} W_{0} ] [ H^{2} + 2 v H ] +$

$+ (1/2) (\partial H)^{2}  + (1/4) \mu^{2} H^{2}  -$

$- (1/16) \mu^{2} v^{2} [ 1 - 4H^{3} / v^{3} - H^{4} /
 v^{4} ]  ] $.

\vspace{12pt}

The weak boson Higgs mechanism masses, in terms of
$v = v_{+} + v_{-} + v_{0}$, are:

\vspace{12pt}

$(\alpha_{w} / 2) v_{+}^{2} = m_{W_{+}}^{2}$ ;

\vspace{12pt}

$(\alpha_{w} / 2) v_{-}^{2} = m_{W_{-}}^{2}$ ;  and

\vspace{12pt}

$(\alpha_{w} / 2) v_{0}^{2} = m_{W_{+0}}^{2}$,

\vspace{12pt}

with $( v = v_{+} + v_{-} + v_{0} ) =  ((\sqrt{2}) /
\sqrt{\alpha_{w}}) ( m_{W_{+}} +  m_{W_{-}} + m_{W_{0}} )$.

\vspace{12pt}

Then:

\vspace{12pt}

$\int [ F_{w} \wedge \star F_{w}  +$

$+ m_{W_{+}}^{2} W_{+} W_{+} +   m_{W_{-}}^{2} W_{-} W_{-} +
  m_{W_{0}}^{2} W_{0} W_{0}  +$

$+ (1/2) [ \alpha_{w} \overline{W_{+}} W_{+} +
  \alpha_{w} \overline{W_{-}} W_{-} +
  \alpha_{w} \overline{W_{0}} W_{0} ] [ H^{2} + 2vH ] +$

$+ (1/2)(\partial H)^{2} + (1/2)(\mu^{2} / 2)H^{2} -$

$- (1/16 \mu^{2} v^{2} [1 - 4H^{3} / v^{3} -
H^{4} / v^{4}]$.

\vspace{12pt}

\subsection{Higgs Mass}

\vspace{12pt}

The Higgs vacuum expectation value
$v = ( v_{+} + v_{-} + v_{0} )$
is the only particle mass free parameter.

\vspace{12pt}

In the $D_{4}-D_{5}-E_{6}$ model,
$v$ is set so that the electron mass $m_{e} = 0.5110 MeV$.

\vspace{12pt}

Therefore,
$(\sqrt{\alpha_{w}}) / \sqrt{2}) v = m_{W_{+}} +
m_{W_{-}} +  m_{W_{0}} = 260.774 GeV$,

\vspace{12pt}

the value chosen so that the electron mass
(which is to be determined from it) will be 0.5110 MeV.

\vspace{12pt}

In the $D_{4}-D_{5}-E_{6}$ model, $\alpha_{w}$ is calculated
to be  $\alpha_{w} = 0.2534577$,

so $\sqrt{\alpha_{w}}$ = 0.5034458 and $v$ = 732.53 GeV.

\vspace{12pt}

The Higgs mass $m_{H}$ is given by the term

$(1/2)(\partial H)^{2} - (1/2)(\mu^{2} / 2)H^{2} =$
$ (1/2) [ (\partial H)^{2} -  (\mu^{2}/2) H^{2} ] $

to be $m_{H}^{2}  =  \mu^{2} / 2  =  \lambda v^{2} / 2$,
 so that $m_{H} = \sqrt{(\mu^{2} / 2)} =
 \sqrt{\lambda} v^{2} / 2)$ .

\vspace{12pt}

$\lambda$ is the scalar self-interaction strength.  It
should be the product of the "weak charges" of two
scalars coming from the reduced 4 dimensions in $Spin(4)$,
which should be the same as the weak charge of the surviving
weak force $SU(2)$ and therefore just the square of the
$SU(2)$ weak charge, $\sqrt{(\alpha_{w}^{2})} = \alpha_{w}$,
where $\alpha_{w}$ is the $SU(2)$ geometric force strength.

Therefore $\lambda = \alpha_{w}  = 0.2534576$,
$\sqrt{\lambda} = 0.5034458$, and $v$ = 732.53 GeV,

so that the mass of the Higgs scalar is
$m_{H} = v \sqrt(\lambda / 2)$  = 260.774 GeV.

\newpage

\section{Spinor Fermion part of the Lagrangian}

Consider the spinor fermion term $\int {\overline{S_{8\pm}}
\not \!  \partial_{8} S_{8\pm}}$

\vspace{12pt}

For each of the surviving 4-dimensional $4$ and reduced
4-dimensional $\perp 4$ of 8-dimensional spacetime,
the part of $S_{8\pm}$ on which the Higgs $SU(2)$ acts
locally is $Q_{3} = {\bf{R}}P^{1} \times S^{2}$.

\vspace{12pt}

It is the Silov boundary of the bounded domain $D_{3}$
that is isomorphic to the symmetric space
$\overline{D_{3}} = Spin(5)/SU(2) \times U(1)$.

\vspace{12pt}

The Dirac operator
$\not \!  \partial_{8} $ decomposes as
$\not \!  \partial = \not \!  \partial_{4}  +
\not \!  \partial_{\perp 4}$,
where

$\not \!  \partial_{4}$ is the Dirac operator
corresponding to the surviving spacetime $4$ and

$\not \!  \partial_{\perp 4}$ is the Dirac operator
corresponding to the reduced 4 $\perp 4$.

\vspace{12pt}

Then the spinor term is
  $\int {\overline{S_{8\pm}} \not \!  \partial_{4} S_{8\pm}} +
\overline{S_{8\pm}} \not \!  \partial_{\perp 4} S_{8\pm}$

The Dirac operator term   $\not \!  \partial_{\perp 4}$
in the reduced $\perp 4$ has dimension of mass.

\vspace{12pt}

After integration
$\int {\overline{S_{8\pm}} \not \!
\partial_{\perp 4} S_{8\pm}}$
over the reduced $\perp 4$,

$\not \! \partial_{\perp 4}$ becomes the real scalar
Higgs scalar field $Y = (v + H)$ that comes

from the complex $SU(2)$ doublet $\Phi$ after
action of the Higgs mechanism.

\vspace{12pt}

If integration over the reduced $\perp 4$ involving
two fermion terms $\overline{S_{8\pm}}$ and $S_{8\pm}$
is taken to change the sign by $i^{2} = -1$, then,
by the Higgs mechanism,

$\int \overline{S_{8\pm}} \not \! \partial_{\perp 4} S_{8\pm}
\rightarrow  \int(\int_{\perp 4}   \overline{S_{8\pm}}
\not \!  \partial_{\perp 4} S_{8\pm} )  \rightarrow  $

$\rightarrow - \int   \overline{S_{8\pm }} YY S_{8\pm }  =
- \int   \overline{S_{8\pm }} Y(v + H)  S_{8\pm }$,

\vspace{12pt}

where:

$H$ is the real physical Higgs scalar,
$m_{H} = v \sqrt(\lambda / 2)$  = 261 GeV, and
$v$ is the vacuum expectation value of the scalar field $Y$,
the free parameter in the theory that sets the mass scale.

\vspace{12pt}

Denote the sum of the three weak boson masses by
$\Sigma_{m_{W}}$.

\vspace{12pt}

$v = \Sigma_{m_{W}}((\sqrt{2}) / \sqrt{\alpha_{w}}) =
260.774 \times \sqrt{2} / 0.5034458 = 732.53 GeV$,

a value chosen so that the electron mass will be 0.5110 MeV.

\vspace{12pt}

\subsection{Yukawa Coupling and Fermion Masses}

\vspace{12pt}

$Y$ is the Yukawa coupling between fermions and
the Higgs field.

\vspace{12pt}

$Y$ acts on all 28 elements

(2 helicity states for each of the 7 Dirac particles
and 7 Dirac antiparticles) of the Dirac fermions in
a given generation, because all of them are in the
same Spin(8) spinor representation.

\vspace{12pt}

Denote the sum of the first generation Dirac fermion masses
by $\Sigma_{f_{1}}$.

\vspace{12pt}

Then $Y = (\sqrt{2}) \Sigma_{f_{1}} / v$, just as
$\sqrt(\alpha_{w}) = (\sqrt{2}) \Sigma_{m_{W}} / v$.

\vspace{12pt}

$Y$ should be the product of two factors:

\vspace{12pt}

$e^{2}$, the square of the electromagnetic charge
$e = \sqrt{\alpha_{E}}$ , because in the term
$\int(\int_{\perp 4} \overline{S_{8\pm }} \not \!
\partial _{\perp 4} S_{8\pm } )  \rightarrow
 - \int \overline{S_{8\pm }} Y(v + H)  S_{8\pm }$
each of the Dirac fermions $S_{8\pm}$ carries
electromagnetic charge proportional to $e$ ; and

\vspace{12pt}

$1/g_{w}$, the reciprocal of the weak
charge $g_{w} = \sqrt{\alpha_{w}}$,
because an $SU(2)$ force, the Higgs $SU(2)$, couples
the scalar field to the fermions.

\vspace{12pt}

Therefore $\Sigma_{f_{1}} = Y v / \sqrt{2} =
 (e^{2} / g_{w}) v / \sqrt{2}$   = 7.508 GeV and

\vspace{12pt}

$\Sigma_{f_{1}} / \Sigma_{m_{W}} = (e^{2} / g_{w}) v /
g_{w} v = e^{2} / g_{w}^{2} = \alpha_{E} / \alpha_{w}$.

\vspace{12pt}

The Higgs term $- \int \overline{S_{8\pm}} Y(v + H)$
$S_{8\pm} = - \int \overline{S_{8\pm}}$
$Yv  S_{8\pm} - \int \overline{S_{8\pm}} YH S_{8\pm } = $
\
$= - int \overline{S_{8\pm}} (\sqrt{2} \Sigma_{f_{1}})
S_{8\pm }  - \int \overline{S_{8\pm}} (\sqrt{2}
\Sigma_{f_{1}} / v)  S_{8\pm}$.

\vspace{12pt}

The resulting spinor term is of the form
$\int  [ \overline{S_{8\pm}} (\not \!  \partial - Yv)
S_{8\pm}    -  \overline{S_{8\pm}} YH S_{8\pm}  ]$ ,

where $(\not \!  \partial - Yv)$ is a
massive Dirac operator.

\vspace{12pt}

How much of the total mass  $\Sigma_{f_{1}} =
Y v / \sqrt{2} = 7.5 GeV$ is allocated to each of the first
generation Dirac fermions is determined by calculating
the individual fermion masses in the $D_{4}-D_{5}-E_{6}$
model, and

\vspace{12pt}

those calculations also give the values of

$\Sigma_{f_{2}} = 32.9 GeV$,
$\Sigma_{f_{3}} = 1,629 GeV$, and

individual second and third generation fermion masses.

\vspace{12pt}

The individual tree-level lepton masses and quark constituent
masses \cite{SM1} are:

$m_{e}$ = 0.5110 MeV (assumed);

$m_{\nu_{e}}$ = $m_{\nu_{\mu}}$ = $m_{\nu_{\tau}}$ = 0;

$m_{d}$ = $m_{u}$ = 312.8 MeV  (constituent quark mass);

$m_{\mu}$ = 104.8 MeV;

$m_{s}$ =  625 MeV  (constituent quark mass);

$m_{c}$ =  2.09 GeV  (constituent quark mass);

$m_{\tau}$ =  1.88 GeV;

$m_{b}$ =  5.63 GeV  (constituent quark mass);

$m_{t}$ = 130 GeV  (constituent quark mass).

\vspace{12pt}

The following formulas use the above masses to
calculate Kobayashi-Maskawa parameters:

phase angle $\epsilon = \pi / 2$

$\sin{\alpha} = [m_{e}+3m_{d}+3m_{u}] /
\sqrt{ [m_{e}^{2}+3m_{d}^{2}+3m_{u}^{2}] +
[m_{\mu}^{2}+3m_{s}^{2}+3m_{c}^{2}] }$

$\sin{\beta} = [m_{e}+3m_{d}+3m_{u}] /
\sqrt{ [m_{e}^{2}+3m_{d}^{2}+3m_{u}^{2}] +
[m_{\tau}^{2}+3m_{b}^{2}+3m_{t}^{2}] }$

$\sin{\tilde{\gamma}} = [m_{\mu}+3m_{s}+3m_{c}] /
\sqrt{ [m_{\tau}^{2}+3m_{b}^{2}+3m_{t}^{2}] +
[m_{\mu}^{2}+3m_{s}^{2}+3m_{c}^{2}] }$

$\sin{\gamma} = \sin{\tilde{\gamma}}
\sqrt{\Sigma_{f_{2}} / \Sigma_{f_{1}}}$

\vspace{12pt}

The resulting Kobayashi-Maskawa parameters are:

\[
\begin{array}{|c|c|c|c|}
\hline
& d & s & b
  \\
\hline
u & 0.975 & 0.222 & -0.00461 i  \\
c & -0.222 -0.000191 i & 0.974 -0.0000434 i & 0.0423  \\
t & 0.00941 -0.00449 i & -0.0413 -0.00102 i & 0.999  \\
\hline
\end{array}
\]

\newpage

\section{Parity Violation, W-Boson Masses, and $\theta_{w}$}

In the $D_{4}-D_{5}-E_{6}$ model prior to dimensional
reduction, the fermion particles are all massless
at tree level.

The neutrinos obey the Weyl equation and must remain
massless and left-handed at tree level after dimensional
reduction.

The electrons and quarks obey the Dirac equation and
acquire mass after dimensional reduction.

After dimensional reduction, the charged $W_{\pm}$ of
the $SU(2)$ weak force can interchange Weyl fermion
neutrinos with Dirac fermion electrons.

\vspace{12pt}

\subsection{Massless Neutrinos and Parity Violation}

\vspace{12pt}

It is required (as an ansatz or part of the
$D_{4}-D_{5}-E_{6}$ model)

that the charged $W_{\pm}$ neutrino-electron interchange
must be symmetric

with the electron-neutrino interchange,
so that the absence of

right-handed neutrino particles requires that
the charged $W_{\pm}$ $SU(2)$

weak bosons act only on left-handed electrons.

\vspace{12pt}

It is also required (as an ansatz or part of the
$D_{4}-D_{5}-E_{6}$ model) that each gauge boson must
act consistently on the entire Dirac fermion particle
sector, so that the charged $W_{\pm}$ $SU(2)$ weak bosons
act only on left-handed fermions of all types.

\vspace{12pt}

Therefore, for the charged $W_{\pm}$ $SU(2)$ weak bosons,
the 4-dimensional spinor fields $S_{8\pm}$ contain only
left-handed particles and right-handed antiparticles.

So, for the charged $W_{\pm}$ $SU(2)$ weak bosons,
$S_{8\pm}$ can be denoted $S_{8 \pm L}$.

\vspace{12pt}

\subsection{$W_{0}$, $Z$, and $\theta_{w}$}

\vspace{12pt}

The neutral $W_{0}$ weak bosons do not interchange Weyl
neutrinos with Dirac fermions, and so may not entirely
be restricted to left-handed spinor particle fields
$S_{8\pm L}$, but may have a component that acts on
the full right-handed and left-handed spinor particle
fields $S_{8\pm} = S_{8\pm L} + S_{8\pm R}$.

\vspace{12pt}

However, the neutral $W_{0}$ weak bosons are related to
the charged $W_{\pm}$ weak bosons by custodial $SU(2)$
symmetry, so that the left-handed component of the
neutral $W_{0}$ must be equal to the left-handed (entire)
component of the charged $W_{\pm}$.

\vspace{12pt}

Since the mass of the $W_{0}$ is greater than the mass
of the $W_{\pm}$, there remains for the $W_{0}$ a component
acting on the full $S_{8\pm} = S_{8\pm L} + S_{8\pm R}$
spinor particle fields.

\vspace{12pt}

Therefore the full $W_{0}$ neutral weak boson interaction
is proportional to
$(m_{W_{\pm}}^{2} / m_{W_{0}}^{2})$ acting on $S_{8\pm L}$
and
$(1 - (m_{W_{\pm}}^{2} / m_{W_{0}}^{2}))$ acting

on $S_{8\pm} = S_{8\pm L} + S_{8\pm R}$.

\vspace{12pt}

If $(1 - (m_{W_{\pm}}2 / m_{W_{0}}^{2}))$ is defined to be
$\sin{\theta_{w}}^{2}$ and denoted by $\xi$, and

\vspace{12pt}

if the strength of the $W_{\pm}$ charged weak force
(and of the custodial $SU(2)$ symmetry) is denoted by $T$,

\vspace{12pt}

then the $W_{0}$ neutral weak interaction can be written as:

$W_{0L} \sim T + \xi$ and $W_{0R} \sim \xi$.

\vspace{12pt}

The $D_{4}-D_{5}-E_{6}$ model allows calculation of
the Weinberg angle $\theta_{w}$, by
$m_{W_{+}} = m_{W_{-}} = m_{W_{0}} \cos{\theta_{w}}$.

\vspace{12pt}

The Hopf fibration of $S^{3}$ as

$S^{1} \rightarrow  S^{3} \rightarrow  S^{2}$

gives a decomposition of the $W$ bosons
into the neutral $W_{0}$ corresponding to $S^{1}$ and
the charged pair $W_{+}$ and $W_{-}$ corresponding
to $S^{2}$.

\vspace{12pt}

The mass ratio of the sum of the masses of
$W_{+}$ and $W_{-}$ to
the mass of $W_{0}$
should be the volume ratio of
the $S^{2}$ in $S^{3}$ to
the $S^{1}$ in ${S3}$.

\vspace{12pt}

The unit sphere $S^{3} \subset R^{4}$ is
normalized by $1 / $2.

\vspace{12pt}

The unit sphere $S^{2} \subset R^{3}$ is
normalized by $1 / \sqrt{3}$.

\vspace{12pt}

The unit sphere $S^{1} \subset R^{2}$ is
normalized by $1 / \sqrt{2}$.

\vspace{12pt}

The ratio of the sum of the $W_{+}$ and $W_{-}$ masses to
the $W_{0}$ mass should then be
$(2  / \sqrt{3}) V(S^{2}) / (2 / \sqrt{2}) V(S^{1}) =
1.632993$.

\vspace{12pt}

The sum
$\Sigma_{m_{W}} = m_{W_{+}} + m_{W_{-}} + m_{W_{0}}$
has been calculated to be
$v \sqrt{\alpha_{w}} = 517.798  \sqrt{0.2534577} =
260.774 GeV$.

\vspace{12pt}

Therefore,
$\cos{\theta_{w}}^{2} = m_{W_{\pm}}^{2 } /
m_{W_{0}}^{2} = (1.632993/2)^{2} = 0.667$ , and
$\sin{\theta_{w}}^{2} = 0.333$,
so $m_{W_{+}} = m_{W_{-}} = 80.9 GeV$, and
$m_{W_{0}} = 98.9 GeV$.

\vspace{12pt}

\subsection{Corrections for $m_{Z}$ and $\theta_{w}$}

\vspace{12pt}

The above values must be corrected for the fact that
only part of the $w_{0}$ acts through the
parity violating $SU(2)$ weak force and the rest
acts through a parity conserving $U(1)$
electromagnetic type force.

\vspace{12pt}

In the $D_{4}-D_{5}-E_{6}$ model, the weak
parity conserving $U(1)$ electromagnetic type force
acts through the $U(1)$ subgroup of $SU(2)$,
which is not exactly like the $D_{4}-D_{5}-E_{6}$
electromagnetic $U(1)$ with force strength
$\alpha_{E} = 1 / 137.03608 = e^{2}$.

\vspace{12pt}

The $W_{0}$ mass $m_{W_{0}}$ has two parts:

the parity violating $SU(2)$ part $m_{W_{0\pm}}$ that is
equal to $m_{W_{\pm}}$ ; and

the parity conserving part $m_{W_{00}}$ that acts like a
heavy photon.

\vspace{12pt}

As $m_{W_{0}}$ = 98.9 GeV = $m_{W_{0\pm}} + m_{W_{00}}$, and
as $m_{W_{0\pm}} = m_{W_{\pm}} = 80.9 GeV$,

we have $m_{W_{00}} = 18 GeV$.

\vspace{12pt}

Denote by $\tilde{\alpha_{E}} = \tilde{e}^{2}$ the force
strength of the weak parity conserving $U(1)$
electromagnetic type force that acts through the
$U(1)$ subgroup of $SU(2)$.

\vspace{12pt}

The $D_{4}-D_{5}-E_{6}$ electromagnetic force strength
$\alpha_{E} = e^{2} = 1 / 137.03608$ was calculated using
the volume $V(S^{1})$ of an $S^{1} \subset R^{2}$,
normalized by $1 / \sqrt{2}$.

\vspace{12pt}

The $\tilde{\alpha_{E}}$ force is part of the $SU(2)$ weak
force whose strength $\alpha_{w} = w^{2}$ was calculated
using the volume $V(S^{2}) of an S^{2} \subset  R^{3}$,
normalized by $1  / \sqrt{3}$.

\vspace{12pt}

Also, the $D_{4}-D_{5}-E_{6}$ electromagnetic force
strength $\alpha_{E} = e^{2}$ was calculated using a
4-dimensional spacetime with global structure of
the 4-torus $T^{4}$ made up of four $S^{1}$ 1-spheres,

while the $SU(2)$ weak force strength
$\alpha_{w} = w^{2}$ was calculated using two 2-spheres
$S^{2} \times S^{2}$, each of which contains one 1-sphere of
the $\tilde{\alpha_{E}}$ force.

\vspace{12pt}

Therefore
$\tilde{\alpha_{E}} = \alpha_{E} (\sqrt{2} /
\sqrt{3})(2 / 4) = \alpha_{E} / \sqrt{6}$,
$\tilde{E}  = e / 4 \sqrt{6} = e / 1.565$ , and

the mass $m_{W_{00}}$ must be reduced to an effective value

$m_{W_{00}eff} = m_{W_{00}} / 1.565$ = 18/1.565 =
11.5 GeV

for the $\tilde{\alpha_{E}}$ force to act like
an electromagnetic force in the 4-dimensional

spacetime of the $D_{4}-D_{5}-E_{6}$ model:

$\tilde{E} m_{W_{00}} = e (1/5.65) m_{W_{00}} = e m_{Z_{0}}$,

\vspace{12pt}

where the physical effective neutral weak boson is
denoted by $Z$ rather than $W_{0}$.

\vspace{12pt}

Therefore, the correct $D_{4}-D_{5}-E_{6}$ values for
weak boson masses and the Weinberg angle are:

\vspace{12pt}

$m_{W_{+}} = m_{W_{-}} = 80.9 GeV$;

\vspace{12pt}

$m_{Z} = 80.9 +11.5 = 92.4 GeV$; and

\vspace{12pt}

$\sin{\theta_{w}}^{2} = 1 - (m_{W_{\pm}} /
m_{Z})^{2} = 1 - 6544.81/8537.76 = 0.233$.

\vspace{12pt}

Radiative corrections are not taken into account here,
and may change the $D_{4}-D_{5}-E_{6}$ value somewhat.

\vspace{12pt}

\newpage

\section{Some Errata for Previous Papers}

hep-th/9402003 \cite{SM4} had a typographical error of
only $S_{8+}$ or $S_{8-}$ instead of $S_{8\pm}$.
The correct 8-dim Lagrangian is:

$$
\int_{V_{8}} F_{8} \wedge \star F_{8} +
\partial_{8}^{2} \Phi_{8} \star \partial_{8}^{2} \Phi_{8} +
\overline{S_{8\pm}} \not \! \partial_{8} S_{8\pm} +
GF + GH
$$

\vspace{12pt}

hep-th/9302030 \cite{SM2} had erroneously
used $\times$ instead of $\oplus$ for the
fermion spinor space.
The correct full fermion space of first generation
particles and antiparticles is

$S_{8+} \oplus S_{8\pm} = ({\bf{R}}P^{1} \times S^{7})
\oplus ({\bf{R}}P^{1} \times S^{7})$.

It is the Silov boundary of the 32(real)-dimensional
bounded complex domain corresponding to the $Type V$ HJTS
$E_{6}/(Spin(10) \times U(1)$ \cite{SM2}

\vspace{12pt}

hep-th/9302030 \cite{SM2} had erroneously listed
$SU(3)/SU(2) \times U(1)$, instead of
$Spin(5)/SU(2) \times U(1) = Spin(5)/Spin(3) \times U(1)$,
as the $Type IV_{3}$ HJTS corresponding to the
6(real)-dimensional bounded complex domain on whose
Silov boundary the gauge group $SU(2)$ naturally acts.

The corrected table is:

The $Q$ and $D$ manifolds for the gauge groups of
the four forces are:

\[
\begin{array}{|c|c|c|c|c|}
\hline
Gauge & Hermitian & Type & m
& Q  \\
Group & Symmetric & of & & \\
& Space & D & & \\
\hline
& & & & \\
Spin(5) & Spin(7) \over {Spin(5) \times U(1)}
& IV_{5} &4 & {\bf R}P^1 \times S^4 \\
& & & & \\
SU(3) & SU(4) \over {SU(3) \times U(1)}
& B^6 \: (ball) &4 & S^5 \\
& & & & \\
SU(2) & Spin(5) \over {SU(2) \times U(1)}
& IV_{3} & 2 & {\bf R}P^1 \times S^2 \\
& & & & \\
U(1) & -  & - & 1  & - \\
& & & & \\
\hline
\end{array}
\]


\vspace{12pt}

\newpage

\begin{thebibliography}{99}

\bibitem{BP} V. Barger and R. Phillips,
{\it Collider Physics}, Addison-Wesley (1987).

\bibitem{GS} M. G\"{o}ckeler and T. Sch\"{u}cker,
{\it Differential geometry, gauge theories, and gravity},
Cambridge (1987).

\bibitem{KN} S. Kobayashi and K. Nomizu,
{\it Foundations of Differential Geometry,
Volume I}, Wiley (1963).

\bibitem{McM} S. MacDowell and F. Mansouri,
Phys. Rev. Lett. {\bf 38}
(1977) 739.

\bibitem{MM} M. Mayer,  {\it The Geometry of
Symmetry Breaking in Gauge Theories},  Acta
Physica Austriaca, Suppl. XXIII (1981) 477-490.

\bibitem{NOS} J. Nieto, O. Obreg\'{o}n, and J.
Socorro, {\it The gauge theory of the de-Sitter
group and Ashtekar formulation}, preprint:
gr-qc/9402029.

\bibitem{LO} L. O'Raifeartaigh, {\it Group structure of
gauge theories}, Cambridge (1986).

\bibitem{SM1} F. Smith,  {\it Calculation of 130 GeV Mass
for T-Quark}, preprint: THEP-93-2; hep-ph/9301210;
clf-alg/smit-93-01(1993).


\bibitem{SM2} F. Smith, {\it Hermitian Jordan
Triple Systems, the Standard Model plus Gravity,
and $\alpha_E$ = $1/137.03608$},
preprint: THEP-93-3; hep-th/9302030.

\bibitem{SM3} F. Smith, {\it Sets and $\bf C^{n}$;
Quivers and $A-D-E$;
Triality; Generalized Supersymmetry; and
$D_{4} - D_{5} - E_{6}$},
preprint:  THEP-93-5; hep-th/9306011.

\bibitem{SM4} F. Smith,  {\it $SU(3)
\times SU(2) \times U(1)$, Higgs, and Gravity
from $Spin(0,8)$ Clifford Algebra $Cl(0,8)$},
preprint: THEP-94-2; hep-th/9402003 (1994).

\end{thebibliography}
\end{document}